# The Danger of a Big Data Episteme and the Need to Evolve GIS

Sean P. Gorman – Timbr LLC, Charlottesville, VA USA

## Abstract

The emergence of "Big Data" as a dominant technology meme challenges Geography's technical underpinnings, found in GIS, while engaging the discipline in a conversation about the meme's impact on society. This allows scholars to engage collaboratively from both a computationally quantitative and critically qualitative perspective. For Geography there is an opportunity to point out these shortcomings through critical appraisals of "Big Data" and its reflection of society. Complimentarily this opens the door to developing methodologies that will allow for a more realistic interpretation of "Big Data" analysis in the context of an unfiltered societal view.

## Big Data as a Meme

"Big Data" is a popular technological meme that has become pervasive in the language discussing a variety of computing challenges and trends. The term "Big Data" has several characteristics often associated with popular technology memes:

1) The component words "big" and "data" are both broad and general
2) The term is open to multiple interpretations
3) The words can easily be composited with other terms to further spread the meme – Big Science, Big Complexity, Big Privacy

"Big Data" is similar in trajectory to "open source" and "web 2.0" memes that lead to a plethora of "open" prefixes and "2.0" suffixes. While the semantics of the creation and use of the term "Big Data" is a fascinating road to walk, this position paper will focus more on the methodological than the critical issues of the meme. Specifically, the paper will examine the role of geography in "Big Data" through the challenges it creates for computation, methodology, and interpretation. Further, it will explore the impact of "Big Data" on the discipline of Geography as seen through the lens of GIS. It should be kept in mind that the impacts of "Big Data" go beyond just quantitative approaches. They may also impact qualitative research as seen through the emerging works in "digital humanities".

While it's difficult to pin down a general definition of "Big Data" it is useful to have a starting point for understanding the role of Geography today and in the future. The short definition of "Big Data" is that it encompasses "a collection of data sets so large and complex that it becomes difficult to process using on-hand database management tools or traditional data processing applications (Wikipedia 2013)." Further, the unique

characteristics of these data sets that make them difficult to manage with traditional tools are:

- Volume – the size of data that must be managed
- Velocity – the speed at which that data arrives and needs to be processed/analyzed
- Variety – the types of data handled include structured and unstructured data (i.e. text, sensor output, GPS, video, audio, log files etc.)
- Veracity – the accuracy and precision of data is variable

In addressing the role of geography in "Big Data" one of the key takeaways is that not only is geography just one type, of several types of data, it is also often inconsistent across a single data set. For instance, in many mobile and social applications users decide whether to include their location or not. This is indicative of a larger trend seen in "location as a feature".

**The Emergence of Location as Feature**

One of the important producers of "Big Data" has been the growth of mobile, social and location applications, often called SoLoMo (Meeker 2011). While geographic information sciences has largely evolved along a path of increasing specialization and complexity driven by professionals, SoLoMo has emerged as a general technology trend quickly making location ubiquitous, driven by consumers. By its very nature SoLoMo has been centered on self service, and allowing the consumer masses toseamlessly interact with location and geography. This technology shift has been driven by several evolving factors and events. First, GPS enabled a larger number of people to create geographic data. This was followed by the incorporation of GPS into commodity technologies like mobile devices. In addition location has permeated up the information technology stack with W3C specifications for adding location to Web browsers, and even the inclusion of location into desktop operating systems. The location component created by these technologies is one data feature of an existing baseline, and not a standalone technology as was developed with GIS. Further, the attributes of data went beyond what the computational underpinnings of GIS was originally constructed for – now integrating unstructured data and temporal attributes both at very large volume and high speeds.

The adoption of "location as a feature" has been massive in scale. The graphic below covers just the adoption of mobile/location technologies to drive social applications:

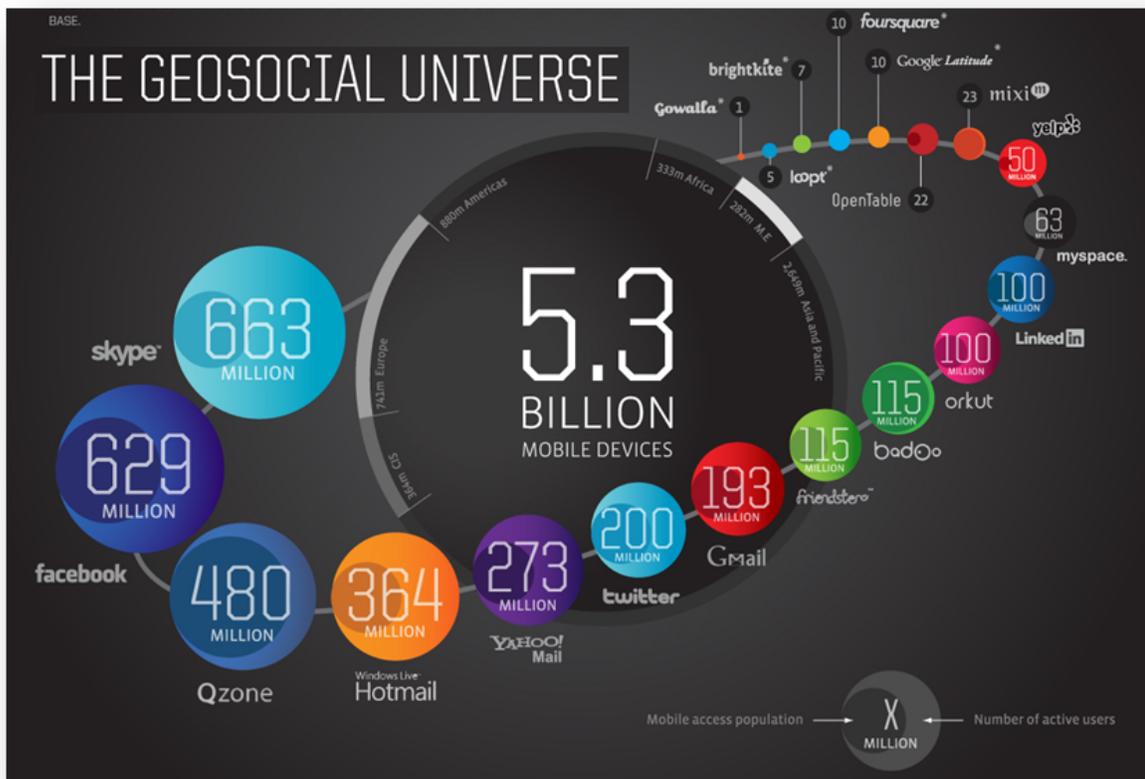

Figure 1: The GeoSocial Universe Adopting Location as a Feature (Jess3 2013)

One of the challenges for Geography as a discipline is that "location as a feature" happened outside the paradigm of geographic information science.  This occurred for several reasons.  1) GIS was built for working with geographic data and location as the center of the operating system.  For the rapidly growing SoLoMo space location was just one of many components that were being leveraged, and was not the center of the operating system.  2) Data flowing from mobile devices through social networks is dynamic and not static.  While time-space has been an active area of study in geographic information science, traditionally data sources have not been unbounded and perpetually updating.  This phenomenon is reinforced by the data characteristics across these emerging services (big events like the Super Bowl and New Years Eve can result in rates of 5,000-6,000 tweets per second from a location) and in massive volumes (155 million tweets in a day) (Twitter 2011).  Put into the larger perspective 90% of all data in the history of humanity has been created in the last two years (Tofel 2011).  This was not the technology paradigm when GIS emerged – data was static, in mostly small volumes and intended for a relatively small audience.  This is not to say GIS has not evolved, but it has iteratively adapted to requirements in its niche of practitioners, and not to the demands of the larger market that is served by "Big Data".

**Monolithic vs. Distributed**

In its inception GIS was predominantly a monolithic application running in a mainframe environment and as it evolved from the command line to the desktop the same monolithic structure persisted. Monolithic systems "help the user carry out a complete task, end to end, and are 'private data silos' rather than parts of a larger system of applications that work together (Wikipedia 2011)." In the case of GIS the complete task was performing geospatial analysis that resulted in an end product to be distributed to viewers. Few other systems at this time generated spatial data, so monolithic structures were not only a popular but also a practical solution. The same structures also dominated word processing and personal finance applications for similar reasons (Wikipedia 2011).

The monolithic structure also matched up well with the philosophical direction of geographic information science. In this construct, GIS was viewed not only as a science but a profession, which required specialty skills and training within geography departments. While this created a corpus of highly trained professionals it also created an insular approach that also manifested itself in the technical architecture of GIS. Data was created, managed, analyzed, visualized and published all within a single system, and the result was considered authoritative and canonical. Operation of the system was by trained professionals. Since data was not created externally a monolithic design was efficient and well suited to the customer set at the time.

As technology has evolved from monolithic to distributed systems, GIS has adapted and evolved as well. First, GIS adapted to a client server environment, and increasingly provides Application Programming Interfaces (API) to data and analysis enabling external consumption. Most recently there have been connections of "Big Data" computational platforms like Hadoop to GIS applications (Github 2013a and Github 2013b). While the hooks into GIS have modernized, the structure has evolved to a more distributed architecture – although user workflows are still geared towards the entire task being done end-to-end in the system.

**Challenges of Scope and Scale**

Computationally "Big Data" and related trends have created a distributed ecosystem with many components and users. As location data comes from an increasing variety of devices and contributors there is a challenge of what mechanism will manage the accuracy and veracity of the data. There are potential problems in both scale and scope applying the current GIS process for determining what is "authoritative data" to these emerging sources of unbounded location data. From a scope perspective it requires the GIS cadre of professionals to be experts in a massive number of subject matter areas – anthropology, sociology, economics, political science, social media, disaster response etc.. Is the disaster response professional on the ground in a better position to determine the quality of data being reported by citizens, or are the GIS professionals back at headquarters?  Should an institution be dependent on having geography/geomatic academic departments generate GIS curriculums to create a new generation of social media analysts before responding to the pressing need to analyze a new source of information?  The inherent problem of having the discipline of geography create a specialty discipline for every aspect of science that has a location or geographic component has long been recognized as "the recurring identity crisis that plagues modern geography and its practitioners (Tuason 1987)".

This is where the problem of scale becomes potentially insurmountable.  The monolithic structure which requires a GIS trained person to dictate data as authoritative has an inherent dependency of requiring a trained person to always be in the loop.  As the volume of location enabled data increases at an exponential rate it raises the real problem of how do the number of GIS professionals scale to keep pace with the speed and volume of the new data that must be verified.  The structure of GIS as a technology and profession was not built to handle massive volumes of external authored data.  Because of its monolithic structure, data was to be generated by professionals solely within the GIS workflow.  Now, Twitter alone is generating millions of location-enabled messages per day.  Simply, there are not enough trained professionals to verify each new piece of data even if they did have the tools.  It is a problem of supply and demand.  The supply of data being generated has far outstripped the supply of trained professionals to verify it - requiring a new paradigm in order to adapt.  This is not to say the concept of verified and unverified data is not critical to effective operations.  It is saying that in order to keep up with the rapidly growing volume of data, the verification and validation of data cannot continue to be purely dependent on trained human professionals doing this by hand or with current tools.  Innovation, automation, statistical inference and the use of crowd sourcing to enable verification and validation of data are greatly needed in order for GIS to successfully adapt.

Issues of privacy and the potential of creating both government and corporate driven surveillance states further complicate this challenge (Dodge and Kitchin 2007).  As humans are taken out of the loop and replaced with algorithmic regulation the application of ethics and governance is unclear.   While this goes outside the scope of this position paper, it is a useful connection of how technological challenges of "Big Data" are directly linked to societal repercussions being focused on by other papers in this journal edition.

**Statistical Challenges of Data at Scale**

The amount of data emerging from "Big Data", where location is one feature of data, is only going to increase at ever-higher velocities.  This new reality is going to require innovative concepts around not only leveraging the crowd for data, but also using the crowd to ascertain the veracity of data.  Traditional concepts like error bounds will fundamentally change because data collection has expanded from just a periodic basis to also include persistent collection from millions of globally distributed sensors.   In this context, error will be a fluid concept and not a static measure.  Metadata needs to also evolve to a fluid concept in these cases.  The requirement for dedicated GIS metadata librarians with hundreds of metadata elements will not scale for "Big Data".  The crowd can be leveraged to verify and update metadata as one potential if not entirely sufficient path.  This has been done with great success for "point of interest" (POI) and road data by projects ranging from Factual to OpenStreetMap respectively.

Further, the concept of sample size and margin of error is being turned upside down.  Previously a small cadre of highly trained professionals made a small number of very precise observations and these were extrapolated to an entire population.  Now, sample sizes come close to the size of the actual population, but are also incredibly biased (i.e. Twitter provides a massive sample but it is biased to only those using Twitter).  Recent work by the Oxford Internet Institute found large biases just in different methods of accessing Twitter to query data for analysis - search API vs. streaming API (Gonzales-Bailon et al 2012).  There is still a lack of fundamental science in understanding what the geographic and demographic biases are of the producers of "Big Data", through the variety of user driven services that create the content.

**The Methodological Challenges of the Variety in Big Data**

The emergence of "location as a feature" in mobile and web apps has not only generated a large amount of new data, but also changed the defining characteristics of the data.  This emerging data is often unstructured, unverified, streaming and unbounded – as noted above this is a different world than the majority of structured GIS data worked with traditionally.

Tackling this data means not only reimagining many current statistical techniques, but also dipping into other disciplines and tool-boxes like natural language processing, statistical mechanics and machine learning to name a few.  Extending Geography to work with these emergent sources of data mean 1) evolving current disciplinary approaches and 2) borrowing from other disciplines to solve new problems.

"Big Data" has several features to it that geographic information science has not commonly focused on, and there is not a solid existing methodological framework for managing.  Challenges in dealing with error, accuracy, and sample bias have been addressed briefly in this paper.  Expanding the list to dealing with the unstructured aspects of big data, unbounded data streams, location as a subset of a larger data set and others goes well beyond the scope of the paper.  It is useful to give a trivial example of how these challenges can make even a simple geographic analysis task challenging though.

Creating thematic maps is one of the most common cartographic outputs and selecting the right classification for data is part of telling the appropriate story of a data analysis.  When the data for a map is static this is a fairly straightforward task.  When data is dynamic and unbounded the task becomes considerably more complex.  Starting with four of the most common approaches to binning data for thematic mapping - equal interval, standard deviation, Jenks natural breaks and quantile – the challenges quickly emerge.  Both standard deviation and Jenk's require the minimum and maximum values of the data distribution to be known.  In the case of an unbounded perpetually updating stream, it is not possible to know what these values will be.  The minimum and maximum values of the data stream historically could be used as a proxy, but these could easily be exceeded in the future causing the mapping to be inaccurate.  Quantile and equal interval can be calculated dynamically since they do not require the bounds of data, but do not cover all data distributions accurately.  Further, these data distributions will change over time so the appropriate binning at time "x" might not also be the correct binning at time "y".  This begins to provide some perspective on the challenges "Big Data" holds for geographic methodologies, which only become more complex when applied to more sophisticated geographic methodologies utilizing "Big Data".

**The Challenge of Interpretation when Big Data Equals the Perceived End of Theory**

The methodological challenges imposed by "Big Data" make interpretation exceeding difficult.  In spite of these obstacles there is a popular conception that "Big Data" will not only be the end of theory (Anderson 2008), but even further the:

> "Belief that big data, harnessed through collective intelligence, would allow us to get at the right answer to every problem, making both representation and deliberation unnecessary" (Morozov 2012).

The panacea aspects of "Big Data" have grown as popular perception, leading to beliefs that results of analyses are applicable to society writ large.  Haklay (2013) has written on the trend of tools and data generated by the technological elite and the biases (Haklay and Budhathoki 2010) in the data generated.  While the concept of a human powered sensor web driven by the adoption of mobile devices is compelling - there is little understanding of the macro-scale dynamics.  Who and who is not connected?  Who contributes and who passively consumes?  How does this breakdown by demographic and geography?  The digital divide is much more than connectivity, but also the participation on the various services riding across networks that generate "Big Data".  What are the "data shadows" created by the interactions of human and machines across networks that compress time and space (Graham 2013)?  The creation of content that feeds "Big Data" both actively and passively has its own distinct geographies and biases that are only beginning to be conceptualized.  Without this parameterization it is incredibly difficult to interpret the results of "Big Data" in the context of global society.

**Conclusion**

This paper began discussing aspects of "Big Data" as a technologic meme.  Exploring how "Big Data" has evolved points to its perceived emergence as an episteme.  What began as an evolution in computation has morphed in popular culture to be a field of scientificity.  Those that work with "Big Data" are even referred to as "data scientists".  The reductionist methods of understanding reality in "Big Data" produce new knowledge and methods for the control of reality.  Yet it is not a reality that reflects the larger society, but instead the small minority contributing content.

For Geography as a discipline there is an opportunity to point out these shortcoming through critical appraisals of "Big Data" and its reflection of society.  Further, there is potentially an even larger opportunity in developing the methodologies that will allow for a more realistic interpretation of "Big Data" analysis in the context of an unfiltered societal view.  To do so the geographic information science aspect of the discipline will need to evolve their approach to data and analysis.   In success this provides a unique opportunity for positivistic and post-positivistic scholars in Geography to collaborate in pushing the discipline forward to an area in need of greater illumination.